\documentclass[seceqn,dvips]{arxstspdf}
\usepackage{dcolumn,lettersp}
\usepackage{shere}
\usepackage{graphicx}
\usepackage{flushend}
\usepackage{stfloats}

\volume{26}
\issue{4}
\pubyear{2011}
\firstpage{502}
\lastpage{516}
\doi{10.1214/09-STS296}

\makeatletter
\newcolumntype{d}[1]{D{.}{.}{#1}}

\newproclaim{comment}{Comment}
\newproclaim{defn}{Definition}[section]

\makeatother

\begin{document}
\begin{frontmatter}

\title{Quick Anomaly Detection by the Newcomb--Benford Law, with
Applications to Electoral Processes Data from the~USA,
Puerto Rico and Venezuela}
\runtitle{Newcomb--Benford Law to Detect Electoral Anomalies}

\begin{aug}
\author[a]{\fnms{Luis} \snm{Pericchi}\corref{}\ead[label=e1]{lrpericchi@uprrp.edu}}\and
\author[b]{\fnms{David} \snm{Torres}\ead[label=e2]{torres.uprrp@gmail.com}}

\runauthor{L. Pericchi and D. Torres}

\affiliation{University of Puerto Rico Rio Piedras}

\address[a]{Luis Pericchi is Professor, Department of Mathematics, College
of Natural Sciences, University of Puerto Rico, Rio Piedras, San Juan PR
00931, USA \printead{e1}.}
\address[b]{David Torres is Ph.D. Student, Department of Mathematics, College of
Natural Sciences, University of Puerto Rico, Rio Piedras, San Juan PR
00931, USA \printead{e2}.}

\end{aug}
%
\begin{abstract}
A simple and quick general test to screen for numerical anomalies is
presented. It can be applied, for example, to electoral processes, both
electronic and manual. It uses vote counts in officially
published voting units, which are typically widely
available and institutionally backed. The test examines the
frequencies of digits on voting counts and rests on the First (NBL1)
and Second Digit Newcomb--Benford Law (NBL2), and in a novel
generalization of the law under restrictions of the maximum number
of voters per unit (RNBL2). We apply the test to the 2004 USA
presidential elections, the Puerto
Rico (1996, 2000 and 2004) governor elections, the 2004 Venezuelan
presidential recall referendum (RRP) and the previous 2000
Venezuelan Presidential election. The NBL2 is compellingly rejected
only in the Venezuelan referendum and only for electronic voting
units. Our original suggestion on the
RRP (Pericchi and Torres, \citeyear{PeriTor04})
was criticized by The Carter Center report (\citeyear{Cartercenter05}). Acknowledging this,
Mebane (\citeyear{Mebane06}) and
The Economist (US) (\citeyear{Economist}) presented voting models and
case studies
in favor of NBL2. Further evidence is presented here.
Moreover, under the RNBL2, Mebane's voting models are valid under wider
conditions. The adequacy of the law is assessed through Bayes
Factors (and corrections of $p$-values) instead of significance
testing, since for large sample sizes
and fixed $\alpha$ levels the null hypothesis is over rejected.
Our tests are extremely simple and can become a standard screening
that a fair electoral process should pass.
\end{abstract}

%

\vspace*{-12pt}
\end{frontmatter}

\section{Introduction}

The Newcomb--Benford Law (NBL) postulates\break that the frequency of
significant digits follow a distribution quite different from the
Uniform (see Tab\-les~\mbox{\ref{T:1}--\ref{T:2}}), as originally discovered by
Newcomb (\citeyear{NewS}) and Benford (\citeyear{Benford}).

Although the NBL works for any
vector of significance numbers, we will use the marginal and joint
distributions of the first or second digits to check the law. Our
goal is to develop methods for initial scrutiny of officially
published electoral data. Official counts (published by the state
electoral authority) are presented in quite variable levels of
aggregation. We call an ``\textit{electoral unit}'' the officially
reported, less aggregated data unit. The composition and size of
these units vary widely in different elections. The data may be
aggregated at county levels (USA) or reported at an elementary polling
unit when no aggregation is performed (Venezuela). If results are
reported from polling machines of around 400 voters or fewer, the
frequency distribution of the first digit of votes counts is heavily
affected. On the other hand, the frequency of second digits should
be less affected. That is why testing the second digit frequency,
although less natural and less powerful than testing the first
digit, is of wider applicability. Our main proposal is to check the
second digit Newcomb--Benford Law NBL2 (also known as 2BL)
or a variation of it by taking into account upper restrictions RNBL2.
However, in cases where the official data is aggregated, as in USA
national electoral data, the first, and even the joint first and second
distribution, fit the data extremely well; see Section~\ref{sec:4}.

The Carter Center was one of the foreign institutions which oversaw
the Venezuelan 2004 Presidential Referendum, and was accepted as a
monitoring external referee by both the government and the \mbox{opposition};
see
\texttt{\href{http://www.cartercenter.org/homepage.html}{http://www.cartercenter.org/}\break
\href{http://www.cartercenter.org/homepage.html}{homepage.html}}.
In the Carter Center Report (\citeyear{Cartercenter05}), pages 132--133, our novel suggestion to
use the Second Digit NBL to scrutinize the Venezuelan 2004
Referendum was criticized on the following 3 grounds: (1) The law is
characteristic of scale invariant data with specific units, like
centimeters or kilograms, so presumably it should not apply to
elections and vote counts. The Newcomb--Benford Law has a simple
justification for numbers which have \textit{units}, like weights,
distances, temperatures, dollars or science constants, on which
scale invariance apply; see, for example,
Pietronero, Tosatti and Vespignani (\citeyear{Piet01}).

However, for unit-less data, like number of votes, a mathematically
well grounded justification exists for using the law. It is based on
a series of now classical contributions by Hill (\citeyear{THill95a,THill96}), that were
summarized in \textit{Statistical Science}.\vadjust{\goodbreak} Hill establishes that NBL holds
asymptotically if the numbers are generated as unbiased mixtures of
different populations, and the more mixing, the better the
approximation. For example, if we generate numbers from a Normal
distribution or from a Cauchy distribution, NBL will be followed
more closely in the latter because the Cauchy distribution is a
scale mixture of Normal distributions. Mixtures of Cauchy
distributions may lead to an even better fit of NBL (Raimi, \citeyear{Raimi}).
Reciprocally, if the NBL is rejected, then the vote counts are
suspect of not being an unbiased realization of numbers sampled from
mixtures of distributions. How to implement this test is the subject
matter of our method. (2) A second criticism was empirical:
``First
digit of precinct-level electoral data for Cook County, the city of
Chicago, and Broward County, Fla. depart significantly from
Benford's Law, primarily because of the relatively constant number
of voters in voting precincts.'' But this criticism is about the
distribution of first digits, and not the distribution of second
digits. For low levels of aggregation of votes, we proposed the
second digit distribution (or a generalization), precisely because
of the limits in the number of voters that produces
``$\ldots$relative constant number of voters in voting precincts.''
The second digit is far less sensitive to constant numbers of voters per polling
unit.\looseness=-1

Compliance with the law based on the first digit is to be expected
only for greater levels of aggregation, as, for example, in the USA
2004 election on which both the first and second digit laws show
impressive fit; see Section~\ref{sec:4.1}. It should also be emphasized that
the results toward NBL are asymptotical in nature, and we require a
substantially large numbers of votes to claim a reasonably
asymptotic situation, which, only perhaps for the Chicago data, can
be claimed among the cases listed by the Carter Center Panel. From
an empirical point of view, in this paper we show several elections
(with larger data sizes) with good fit to NBL (see Section~\ref{sec:4}), where
compliance with the law is the norm rather than the exception. There
is a rapidly increasing number of contributions in which compliance
and violations of the NBL have been presented for electoral votes;
see Pericchi and Torres (\citeyear{PeriTor04}),
Mebane (\citeyear{Mebane06,Mebane07a,Mebane07b}),
Torres et al. (\citeyear{TorresJ2007}) and Buttorff (\citeyear{Buttorff}) among others.
(3) A final criticism, rai\-sed by the panel appointed by the Carter Center, was
that under some (perhaps \textit{over simplistic}) electoral
models,
computer simulations did not yield frequencies of second digits in
accordance with NBL2. The\vadjust{\goodbreak} fact that for some mathematical models
NBL2 is not observed may also be regarded as evidence of the lack
of realism of such models, and more sophisticated idealizations
ought to be searched. In Taylor (\citeyear{Taylor2005,Taylor}) (who was part of
the Carter Center Panel) a very intriguing and brief discussion is
made of the Newcomb--Benford law regarding elections. The claim is
made that the NBL is of ``little use in fraud detection'' for
elections. However, the rationalization covers only the first digit
NBL and not the second digit. Data is simulated from models that can
be criticized for not being realistic, since realistic population
voting models should not be homogeneous on each electoral unit, but
should be mixtures of different populations (see next paragraph).
The claim seems to be that the results of the simulations contradict
NBL for the first, second and third digit laws. However, no measures
of fit are provided, and intriguingly, the figures that cover the
second and third digits have only 9 entries, although there are 10
second and third digits. (See Taylor, \citeyear{Taylor2005}, Figure~8, page 23, Technical
Report version November,~7, 2005). Furthermore, for the second digit
at least, the fit of the votes for and against the government
appear to be markedly different, a fact that is not discussed in the
cited Technical Report.\looseness=-1

The negative criticism of the Carter Center Panel did not convince
everybody. Acknowledging our original suggestion and the Carter
Center Report, Walter Mebane presented an invited conference at the
Annual Meeting of the American Association for the Advancement of
Science which was reported in The Economist (US) (\citeyear{Economist}), on which the
suggestive
term ``\textit{Election Forensics}'' was coined by Mebane. He provides
further support to the use of the 2nd
digit NBL (calling it 2BL) for an initial quick scrutiny of
elections based solely on officially reported data on the current
election and does not require the use of covariates
(Mebane, \citeyear{Mebane06}). Mebane produced simulations from realistic models
of electorate behavior which are consistent with the 2nd digit NBL,
and also presented different types of frauds that are detected by
tests on the 2nd digit NBL (although not all frauds are detected).
His models are an interesting reflection of political behavior,
which are hierarchical mixed population models, denoted here HMPM.
In these models there are two populations of voters at each polling
station: the partisan population strongly in favor of a candidate
and the general population, swinging between candidates. There was,
however, a question about the general applicability of the 2nd digit
Law: Mebane's models produce frequencies\vadjust{\goodbreak} according to NBL2 for some
numbers of voters per unit, say, 2000 or 3000 voters per electoral
unit, but not for others, say, 2250.
We introduced the Restricted Newcomb--Benford Law (RNBL) in Torres N\'{u}\~{n}ez (\citeyear{Torres06}),
before being aware of Mebane's models. It turns out that the RNBL2 is consistent generally
with Mebane's models, which is illustrated in Table~\ref{T:4}.

The NBL1 has been utilized before to check, for example, \textit{tax
fraud} (Nigrini, \citeyear{Nigrini95}), and microarrays data corruption
(Torres N\'{u}\~{n}ez, \citeyear{Torres06}). Its use for elections is timely, since electronic
voting is raising fresh concerns about the possibility of massive
interference with the digital data (Pericchi and Torres, \citeyear{PeriTor04}).

The official electoral data, when not presented with levels of
aggregation, may have a small upper bound, namely, the number of
potential voters. In that respect, when necessary, we proceed in two
ways: (1) Check the second digit number Law instead of the first,
because the second digit is far less affected, if at all by
restrictions on the total; (2) If (1) fails, try the restricted
second digit law RNBL with
realistic upper bounds; see next section. If both fail, then the alarm
is on and further study is required.

The empirical general picture that emerges is that the fit of NBL is
accepted in the elections in USA in Puerto Rico and in the manual
elections in Venezuela. (In USA 2004, even the first digit and the
more complex joint first and second digit test accepts NBL without
restrictions). Electronic voting in Venezuela, in the recall
referendum, however, fails the test and, to some extent, in the
previous presidential elections, adding to the suspicions about
electronic voting, particularly without universal paper checking and
audits, prior to the sending of the data to the central polling
station.

This paper is organized as follows: Section~\ref{sec:2} is
devoted to the description of the law and a generalization. Section~\ref{sec:3}
discusses different methods, alternative to the use of
$p$-values to judge the fit of the models. Section~\ref{sec:4}
presents the data analysis of the USA, Puerto Rico and Venezuelan
elections and Venezuelan recall referendum. Section~\ref{sec:5}
states some conclusions.

%
\begin{table*}
\caption{Newcomb--Benford Law for the first significant digit}\label{T:1}
\begin{tabular*}{\textwidth}{@{\extracolsep{\fill}}lccccccccc@{}}
\hline
\textbf{Digit unit} &$\bolds{1}$ &$\bolds{2}$ &$\bolds{3}$ &$\bolds{4}$ &$\bolds{5}$ &$\bolds{6}$ &$\bolds{7}$ &$\bolds{8}$ &$\bolds{9}$ \\
\hline
Probability
&0.301 &0.176 &0.125 &0.097 &0.079 &0.067 &0.058 &0.051 &0.046
\\
\hline
\end{tabular*}
\end{table*}

%
\begin{table*}[b]
\caption{Newcomb--Benford Law for the second significant digit}\label{T:2}
\begin{tabular*}{\textwidth}{@{\extracolsep{\fill}}lcccccccccc@{}}
\hline
\textbf{Digit unit}&$\bolds{0}$ &$\bolds{1}$ &$\bolds{2}$ &$\bolds{3}$ &$\bolds{4}$ &$\bolds{5}$ &$\bolds{6}$ &$\bolds{7}$ &$\bolds{8}$ &$\bolds{9}$ \\
\hline
Probability &0.120 &0.114 &0.109 &0.104 &0.100 &0.097 &0.093 &0.090 &0.088 &0.085\\
\hline
\end{tabular*}
\end{table*}

\section{Overview on the Newcomb--Benford Framework}\label{sec:2}

Intuitively, most people assume that in a string of numbers sampled
randomly from some body of data, the first nonzero digit could be
any number from~1 through 9, with\vadjust{\goodbreak} all nine numbers being equally
probable. Empirically, however, it has been found that a~law first
discovered by Newcomb and later popularized by Benford is
ubiquitous.

For the first and second digit Newcomb--Benford Laws we have discrete
probability distribution values presented in Table~\ref{T:1} and
Table~\ref{T:2}, respectively, which
are quite different from the Uniform Distribution.

The most general probabilistic justification of the NBL is in Hill (\citeyear{THill96}).

Hill developed the probability theory that justifies the asymptotic
validity of the law for data such as people counts, which do not
have units like grams or meters.

 The aim here is to use and
generalize the New\-comb--Benford~Law in order to apply it to wider classes
of data sets, particularly arising from elections and to verify
their fit to different sets of data with Bayesian statistical
methods.

The general definition of the Newcomb--Benford Law is stated here, on
base 10, for simplicity. First we introduce the simpler laws for
the first and second significant digits. Let $D_1,D_2,\ldots$ denote
the significant digit functions. For example, $D_2(0.154)=5$ gives
the second significant digit:
\begin{eqnarray}
p^B_1(d_1)&=&\mathit{Prob}(D_1=\mbox{First significant digit}=d_1)\nonumber
\\
&=&\log_{10}(1+1/d_1),\qquad d_1=1,2,\ldots, 9,\nonumber
\\[2pt]
p^B_2(d_2)&=&\mathit{Prob}(D_2=\mbox{Second significant digit}=d_2)\nonumber
\\
&=&\sum_{j=1}^9 \log_{10} \bigl(1+1/(10j + d_2)\bigr),\nonumber
\\
\eqntext{d_2=0,1,\ldots,9.}
\end{eqnarray}
For all positive integers $k$, all $d_1\in{\{1,\ldots,9\}}$ and
$d_j\in{\{0,1,\ldots,9\}}$ for $j=2,\ldots,k$ the joint
Newcomb--Benford distribution is
\begin{eqnarray*}
p^B_{1,\ldots,k}(d_1,\ldots,d_k)&=&\mathit{Prob}(D_1=d_1, \ldots, D_k=d_k)
\\
&=&\log_{10}\biggl[1+\frac{1}{\sum_{i=1}^{k}d_i\cdot10^{(k-i)}}\biggr].
\end{eqnarray*}

In the remainder of this section we postulate the way in which the
N--B Law acts under restrictions, when the number of electors per
electoral unit is restricted to be smaller than a relatively small
and known number $K$. This may be important when official data have
not been aggregated. The notation used in the following discussion
is:
\begin{enumerate}
\item$p_i^B(d_i)$ is the Newcomb--Benford Probability
Distribution for the digit $i$ and number $d_i$. These are presented in
Table~\ref{T:1} and Table~\ref{T:2} for the first ($i=1$) and
second significant digit ($i=2$) respectively.
\item$p_i^C(d_i)$ under the constraint $N \leq K$ is the proportion of
the numbers
with $i$th-digit equals to $d_i$ in the set of numbers that are smaller
or equal to $K$, that is, $p_i^C(d_i)=\frac{\sharp{d_i\leq K}}{K}$,
where $\sharp{d_i \leq K}$ is the cardinality of numbers with
$i$th-digit equal to $d_i$ that are no bigger than $K$;
\item$p^U_i(d_i)$ the proportion of numbers with $i$th-digit equal to
$d_i$ if no constraints were
present.
\end{enumerate}
Note that if there is no restriction, then $p^C =
p^U$. However, if $K=800$, for example, then for the first
significant digit, $p_1^B(d_1=2)=0.176$ (see Table~\ref{T:1}),
$p_1^C(d_1=2)=\frac{111}{800}$ and $p_1^U(d_1=2)=\frac{1}{9}$.

%
\begin{table*}
\caption{NBL for first and second digit with and without an upper restriction of 800}\label{T:3}
\begin{tabular*}{\textwidth}{@{\extracolsep{\fill}}lccccccccd{1.4}c@{}}
\hline
&$\bolds{0}$&$\bolds{1}$&$\bolds{2}$&$\bolds{3}$&$\bolds{4}$&$\bolds{5}$&$\bolds{6}$&$\bolds{7}$&\multicolumn{1}{c}{$\bolds{8}$}&$\bolds{9}$\\
\hline
$\mathit{NB}1$ & &0.301&0.176&0.125&0.097&0.079&0.067&0.058&0.0512&0.046\\
$\mathit{CNB}1_{800}$&&0.330&0.193&0.137&0.106&0.087&0.073&0.064&0.006&0.005
\\[6pt]
$\mathit{NB}2$&0.120&0.114&0.109&0.104&0.100&0.097&0.093&0.090&0.088&0.085\\
$\mathit{CNB}2_{800}$&0.121&0.114&0.109&0.104&0.100&0.097&0.093&0.090&0.087&0.085\\
\hline
\end{tabular*}
\end{table*}

\begin{defn}
The Restricted N--B Law\break (RNBL) distribution is
\begin{equation}\label{bouneq}
\qquad p_i(d_i|N \leq K)
=\frac{p_i^B(d_i){p^C_i(d_i)}/{p_i^U(d_i)}}{\sum_{d_i'}
p_i^B(d'_i){p^C_i(d'_i)}/{p^U_i(d'_i)}}.
\end{equation}
\end{defn}

The heuristics behind the RNBL is as follows: sample from sets of
numbers that
obey NBL, but reject the number if and only if it does not obey the
restriction. Note that if $p_i^C \equiv p_i^U$, then the usual
Newcomb--Benford Law (NBL) is recovered, whether there is a
restriction or not. Take as an example the first digit law. If the
numbers are restricted to be less than or equal to $K=9$, there is
no correction to the NBL. But if $K=15$, say, a substantial
correction applies. Note also that the restricted rule is also valid
for lower bound restrictions of the form $N \geq K$ or even for two
sided restrictions.

For positive numbers, there is a simpler expression for the equation
above in terms of the cardinality of the sets induced by the
restriction. It turns out that $p^U_i(d'_i)=\mathit{constant}$ (the constant
is equal to $1/9$ for the first digit and to $1/10$ for the second
digit). This fact allows to cancel out $p^U_i$ in (\ref{bouneq}).
Now let $\sharp{d_i\leq K}$ be the number of positive numbers less
than or equal to $K$, with the $i$th-significant digit equal to
$d_i$. We may now simplify (\ref{bouneq}) as follows:
\begin{eqnarray*} 
p_i(d_i|N \leq K)
&=&
\frac{p_i^B(d_i){p_i^C(d_i)}/{p_i^U(d_i)}}{\sum_{d_i'}
p_i^B(d_i'){p_i^C(d_i')}/{p_i^U(d_i')}}
\\
&&{}\mbox{canceling $p^U_i=c$}\\
&=& \frac{p_i^B(d_i)p_i^C(d_i)}{\sum_{d_i'} p^B_i(d_i')p_i^C(d_i')} \\
&=& \frac{p_i^B(d_i){\sharp\{d_i\leq K \}}/{K}}{\sum_{d_i'}
p_i^B(d'_i){\sharp\{d_i' \leq K\}}/{K}}
\\
&&{}\mbox{canceling }K\\
&=&\frac{p_i^B(d_i)\sharp\{d_i\leq K \}}{\sum_{d_i'}
p_i^B(d_i')\sharp\{d_i' \leq K\}}.
\end{eqnarray*}

This is a simpler expression easier to calculate.

Mebane (\citeyear{Mebane06,Mebane07a,Mebane07b}) introduced realistic models (HMPM models) of
electoral behavior that produced frequencies consistent with the
NBL2 for some numbers of electors per unit, like 2000, but not for
other such as 2250.

Table~\ref{T:4} displays a large simulation with expected maximum
number of voters of 2250 which shows the second digit RNBL to be
more consistent with
HMPM models than the usual second digit NBL, as anticipated.

%
\begin{table}
\vspace*{3pt}
\tabcolsep=4.5pt
\caption{Table with an upper bound of $N=2250$ voters,
that~illustrates the better fit of the restricted law, over~$m=999$
simulations}\label{T:4}
\vspace*{-3pt}
\begin{tabular*}{\columnwidth}{@{\extracolsep{\fill}}lccccc@{}}
\hline
&$\bolds{m}$
&$\bolds{P(H_0|\mathit{data})}$
&$\bolds{p}$\textbf{-values}
&$\bolds{\underline{P}(H_0|\mathit{data})}$\\
\hline
No restrictions &999&0.9996&0.001&\phantom{00}0.018\\
Restrictions &999&1.0000&0.802& \multicolumn{1}{c}{\hspace*{-8.5pt}$>0.5$} \\\hline
\end{tabular*}
\vspace*{-5pt}
\end{table}

\section{Changing $p$-Values to Null Hypothesis Probabilities}\label{sec:3}

The $p$-value is the probability of getting values of the test
statistic as extreme as or more extreme than the value actually
observed given that the null hypothesis is true. For the first
significant digit, the observed chi-squared statistic
$\chi^2_{\mathit{Observed}}$ is given by
\begin{eqnarray}
&&\quad\hspace*{18pt}\chi^2_{\mathit{Observed}}\nonumber
\\[-8pt]\\[-8pt]
&&\qquad\hspace*{18pt} = \mbox{Sample size}\times \sum_{d=1}^9 \frac{(\mathit{Prob}(D_1=d)- f_d)^2}{\mathit{Prob}(D_1=d)},\nonumber
\end{eqnarray}
where $f_d$ is the proportion observed of the digit $d$ as the first
significant digit. For the second significant digit, $D_2=d \in{\{0,1,\ldots,9\}.}$
This is the basis of a classical test of the null hypothesis which is that
the data follows the Newcomb--Benford Law.
If the null hypothesis\vadjust{\goodbreak} is accepted, the data ``passed'' the test. If not, a sort of
inconsistency has been found which opens the possibility of
manipulation of the data. In the electoral process the null
hypothesis is $H_0{}\dvtx{}$The data is consistent with the
Newcomb--Benford proportions for the second significant digit
(in Table~\ref{T:2}), while the alternative $H_1$ means that there is an
inconsistency with the law. It is important to get a quantification
of the evidence in favor of the Null Hypothesis. In our case, if the
data obeys Newcomb--Benford's Law, then the test offers no basis to
suspect undue
intervention in the electoral process.

There is a well known statistical misunderstanding between the
probability that the null hypothesis is true and the $p$-value. One
general way to calibrate $p$-values is through the Universal Upper
Bound, due to Sellke, Bayarri and Berger (\citeyear{Ber}). For a null hypotheses, $H_0$, we have
\[
P_{val}= \mathit{Prob}(\chi^2_\nu\geq\chi^2_{\mathit{Observed}}|\mbox{ null hypothesis is true}),
\]
where $\nu$ is the degrees of freedom, which is equal to 8 for the
first significant digit and 9 for the second and onward. If the
$p$-value is small
(Ex. $p$-values${}<0.05$ or less), it is assumed, based on
uncritical practice and convention, that there
is a significant result. But the \mbox{$p$-value} is not the
probability that the sample arose from the null hypothesis and,
therefore, it should not be interpreted as a probability. The
usefulness and interpretation of a $p$-value is drastically
affected by the sample size.

A useful way to calibrate a $p$-value, under a Robust
Bayesian perspective, is by using the bound that is found as the
minimum posterior probability of $H_0$ that is obtained by changing
the priors over large classes of priors under the alternative
hypothesis. If a priori we have equal prior probabilities for the
two hypotheses, $P(H_0)=P(H_1)=1/2$, and for $p_{val}< e^{-1}$,
then
\begin{eqnarray}\label{ULB}
&&P(H_0|P_{val})\nonumber
\\[-8pt]\\[-8pt]
&&\quad\ge1/\bigl(1+[-e\cdot p_{val}\cdot\log_{e}(p_{val})]^{-1}\bigr).\nonumber
\end{eqnarray}
A full discussion about this matters
can be found in Sellke, Bayarri and Berger (\citeyear{Ber}).
%
\begin{table}
\tablewidth=6cm
\caption{$p$-values in terms of Hypotheses probabilities}\label{T:5}
\begin{tabular*}{\tablewidth}{@{\extracolsep{\fill}}ld{1.4}@{}}
\hline
$\bolds{p_{\mathit{val}}}$
& \multicolumn{1}{c@{}}{$\bolds{\underline{P}(H_0|\mathit{data})}$} \\\hline
0.05 & 0.29 \\
0.01 & 0.11 \\
0.001 & 0.0184 \\
\hline
\end{tabular*}
\vspace*{-6pt}
\end{table}

It is more appropriate to report the Universal Lo\-wer Bound
(\ref{ULB}) than the $p$-value, with respect to the goodness
of fit test of the proportions in the observed digits versus those
proportions specified by the Newcomb--Benford Law. As we can see in
Table~\ref{T:5}, the correction is quite important. This table shows
how much larger this lower bound is than the $p$-values.\vadjust{\goodbreak} Small
$p$-values (i.e., $p_{val}=0.05$) imply that the posterior
probability of the null hypotheses is at least $0.29$, which is not
very strong evidence to reject a null hypothesis.

However, the lower bound correction does not depend on sample size,
so for large sample sizes it can be very conservative. For a full
correction of $p$-values, a Bayes Factor is needed, with the
corresponding posterior probability of the null hypothesis. Next we
compute a very simple Bayes Factor, based on a Uniform prior.

\subsection{Posterior Probabilities with Uniform Priors}

Let
$\Upsilon_1=[1,2,\ldots,8,9]$ and $\Upsilon_2=[0,1,\ldots,8,9]$.\break
The~elements that may appear when the first digit is observed are
members of $\Upsilon_1$ and if we observe the second digit or
higher, the observations are members of $\Upsilon_2$.
Let
\begin{eqnarray*}
\Omega_0&=&\Biggl\{p_1=p_{0 1},p_2=p_{0 2},\ldots,
\\
&&{}\hspace*{19pt}p_k=p_{0 k}\Big| \sum_{i=1}^k p_{0i}=1 \Biggr\}
\end{eqnarray*}
for $k=1,\ldots,9$
in the case of the first digit and $k=0,1,\ldots,9$ for the second
digit.
Then our hypothesis can be written as
\begin{equation}\label{T2:TEST}
\matrix{
H_0 = \Omega_0, \cr
H_1 = \Omega_0^{'},
}
\end{equation}
where $\Omega_0^{'}$ means the complement of $\Omega_0$. In other
words,
\[
\Omega_0^{'}=\{ p_i \neq p_{0i} \mbox{ for at least one } i\in
\Upsilon\}.
\]
As the simplest objective prior distribution assume an uniform prior
for the values of the $p_i's$, then
\begin{eqnarray}
\pi^u(p_1,p_2,\ldots,p_k)
&=&\mathit{constant}=\Gamma(k)\nonumber
\\[-8pt]\\[-8pt]
&=&(k-1)!,\nonumber
\end{eqnarray}
which is
the correct normalization constant, as it is seen from the well-known
integral $\int_{\Omega}\, dp_1 \cdots dp_{k-1}=1/\Gamma(k)$.

We can write the posterior probability of $H_0$ in terms of the
Bayes Factor. Let $\mathbf{x}$ be the data vector, then the Bayes Factor
is
\begin{equation}
B_{01}=\frac{P(H_0|\mathbf{x})P(H_1)}{P(H_1|\mathbf{x})P(H_0)}.
\end{equation}
If
we have nested models and $P(H_0)=P(H_1)=\frac{1}{2}$, then the
Bayes Factor reduces to
\begin{equation}
B_{01}=\frac{P(H_0|\mathbf{x})}{P(H_1|\mathbf{x})},
\end{equation}
where
\begin{equation}\label{T2:PHOB01} P(H_0|\mathbf{x})
=\frac{B_{01}}{B_{01}+1}.
\end{equation}
For the $i$th significant digit, the
data vector is $ \textbf{n} = (n_1,n_2,\ldots,n_k)$, where $n_d$ is
the frequency with\break which~$d$ is the $i$th significant digit in the
data.
Using the definition of a Bayes Factor with a simple hypothesis, we
have
\begin{eqnarray*}
B_{01}&=& {f(n_1,\ldots,
n_k|\Omega_0)}
\\[2pt]
&&{}\big/\biggl(\int_{\Omega_0^{'}}f(n_1,\ldots,n_k|\Omega_0^{'})
\\[2pt]
&&{}\hspace*{30pt}\cdot\pi^U(p_1,\ldots, p_k)\,dp_1 \cdots \,dp_{k-1}\biggr),
\end{eqnarray*}
with $\sum_{i \in\Upsilon}p_i=1 $ and $p_i \geq0\ \forall i \in\Upsilon$. Substituting our assumptions,
\begin{eqnarray*}
&&\hspace*{-1pt}B_{01}
\\[2pt]
&&\hspace*{-1pt}\quad=
\frac{n!}{\prod_{i=1}^k n_i!}\prod_{i=1}^k p_{i0}^{n_i}
\\[2pt]
&&\hspace*{-1pt}\qquad{}\bigg/\Biggl((k-1)!\int_{-\infty}^{+\infty}\frac{n!}{\prod_{i=1}^kn_i!}
\\[2pt]
&&\qquad\hspace*{76pt}{}\cdot
\prod_{i=1}^k p_{i}^{n_i+1-1}\,dp_1\cdots\,dp_{k-1}\Biggr).
\end{eqnarray*}

After canceling factorial terms and using the identity
\[
\int_{-\infty}^{+\infty}\prod_{i=1}^k p_{i}^{n_i+1-1}\,dp_1\cdots\,dp_k=\frac{\prod_{i=1}^k\Gamma(n_i+1)}{\Gamma(n+k)},
\]
we obtain a simplified expression for $B_{01}$,
\begin{equation}
\hspace*{10pt}B_{01}=\frac{p_{10}^{n_1}p_{20}^{n_2}\cdots p_{k0}^{n_k}}{(k-1)!
{\prod_{i=1}^k\Gamma(n_i+1)}/{\Gamma(n+k)}}.
\end{equation}
To obtain the
posterior probability using the Bayes Factor (using~\ref{T2:PHOB01}) and substituting $B_{01}$, we get
\begin{eqnarray}
&&\hspace*{22pt}P(H_0|\mathbf{x})\nonumber
\\
&&\hspace*{22pt}\quad={p_{10}^{n_1}p_{20}^{n_2}\cdots
p_{k0}^{n_k}}\big/\biggl({(k-1)!
\frac{\prod_{i=1}^k\Gamma(n_i+1)}{\Gamma(n+k)}}\biggr)\nonumber
\\[-8pt]\\[-8pt]
&&\hspace*{22pt}\qquad{}\Big/\biggl({p_{10}^{n_1}p_{20}^{n_2}\cdots p_{k0}^{n_k}}\nonumber
\\
&&\qquad\hspace*{38pt}{}\big/\biggl((k-1)!\frac{\prod_{i=1}^k\Gamma(n_i+1)}{\Gamma(n+k)}\biggr)+1\biggr).\nonumber
\end{eqnarray}
In Torres N\'{u}\~{n}ez (\citeyear{Torres06}), calculations of posterior probabilities with
several other priors and approximations are presented. The
conclusions are similar to those presented here. [See Berger and Pericchi (\citeyear{BP01})
for priors and approximations in Bayesian Models Selection].

%
\begin{table*}
\caption{Summary USA 2004 Elections}\label{T:6}
\begin{tabular*}{\textwidth}{@{\extracolsep{\fill}}lcd{4.1}d{5.1}d{5.1}d{5.1}r@{}}
\hline
\textbf{United States 2004}
&\textbf{Min}
&\multicolumn{1}{c}{\textbf{1st Qu.}}
&\multicolumn{1}{c}{\textbf{Median}}
&\multicolumn{1}{c}{\textbf{Mean}}
&\multicolumn{1}{c}{\textbf{3rd Qu.}}
& \multicolumn{1}{c@{}}{\textbf{Max.}}\\
\hline
Bush votes &2&1816 &5047 &18380 &14130 &1076000\\
Kerry votes &3&973.2&3225.0&16840.0&9156.0&1908000\\
Nader votes &1&13&31&143.7&85&13251\\
\hline
\end{tabular*}
\vspace*{-8pt}
\end{table*}

%
\begin{table*}
\caption{USA 2004 Elections}\label{T:7}
\begin{tabular*}{\textwidth}{@{\extracolsep{\fill}}lcd{4.0}ccd{3.3}@{}}
\hline
\textbf{United States 2004}
&$\bolds{m}$
&\textbf{Median}
&\multicolumn{1}{c}{$\bolds{P(H_0|\mathit{data})}$}
&$\bolds{p}$\textbf{-values}
&\multicolumn{1}{c@{}}{$\bolds{\underline{P}(H_0|\mathit{data})}$}
\\\hline
NB1 Bush votes&4715&3694&1.000&0.003&0.050\\
NB1 Kerry votes&4714&2603&1.000&0.002&0.034\\
NB1 Nader votes&2822&8&1.000&0.833& \multicolumn{1}{c}{\hspace*{-3.5pt}$>$ 0.5} \\[5pt]
NB2 Bush votes&4708&3713&1.000&0.068& 0.331\\
NB2 Kerry votes&4698&2621&1.000&0.651&\multicolumn{1}{c}{\hspace*{-3.5pt}$>$ 0.5} \\
NB2 Nader votes&2271&44&1.000&0.830&\multicolumn{1}{c}{\hspace*{-3.5pt}$>$ 0.5} \\
\hline
\end{tabular*}
\vspace*{-3pt}
\end{table*}

\section{Results and Data Analysis}\label{sec:4}

We illustrate the use of the First and Second digit Newcomb--Benford
Law with data from the 2004 USA elections, three elections in Puerto Rico and
the Presidential Recall referendum in Venezuela and one previous
Presidential election in that country. We denote by NB1 and NB2 the
analysis according to the first and second digit NBL, respectively.
We show in the tables the value $m$ which denotes the number of
electoral units, and the median number of votes for the respective
candidate on the information units. There is wide variation on the
aggregation of the numbers, with the USA case as the most aggregate,
and Venezuela the least aggregate. That is the reason why the first
digit law is obeyed only in the USA, and the fit is remarkable. In
most cases, the second digit law is also obeyed, without the need to
use the restricted NBL. The case in which the NBL2 was
overwhelmingly violated is presented by the Venezuelan Presidential
recall vote. We attempted to mend
it by restricting the Law for various plausible upper bounds, but the
fit did not improve.

\subsection{United States Elections 2004}\label{sec:4.1}

\begin{figure*}[b]
\vspace*{-3pt}
\includegraphics{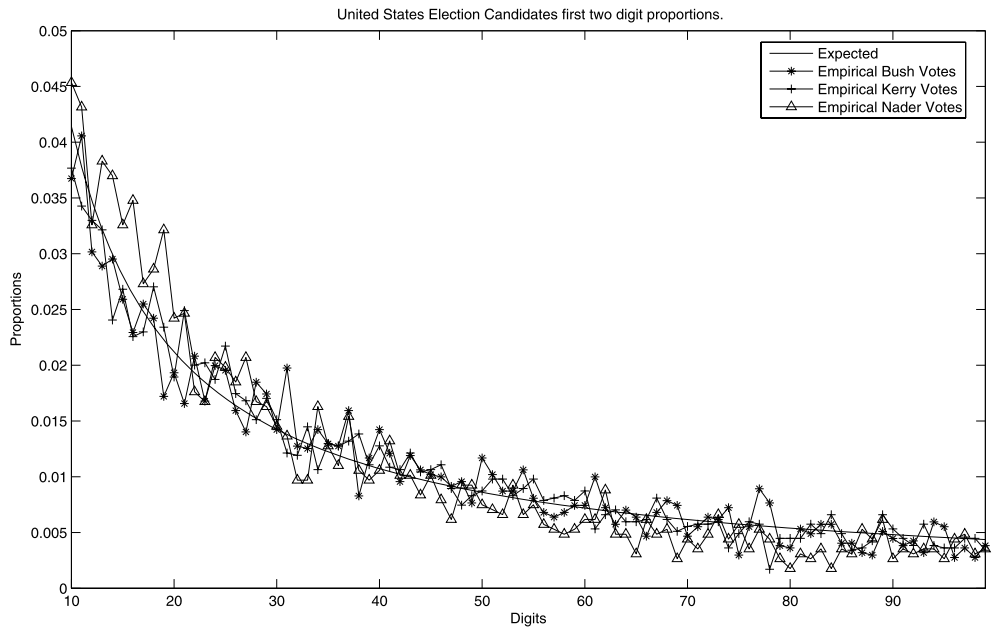}
\caption{Empirical distributions of the first two digits of the
presidential candidates vs. N--B Law for the first two digits.}\label{USAjoint}
\end{figure*}

\begin{figure}

\includegraphics{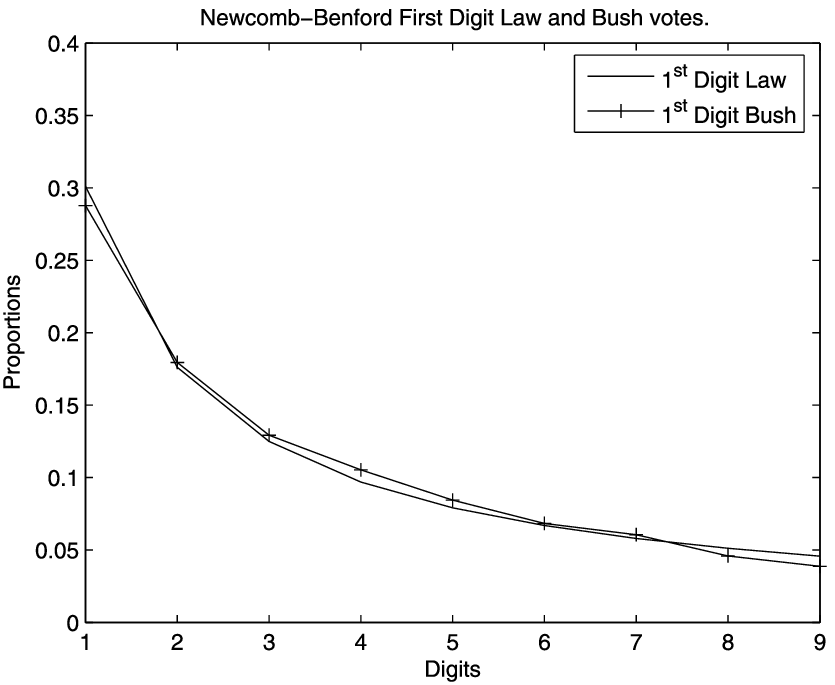}

\caption{Bush's digit proportions vs N--B Law for the $1$st digit.}\label{USABUSHNB1}
\vspace*{-6pt}
\end{figure}

\begin{figure}
\includegraphics{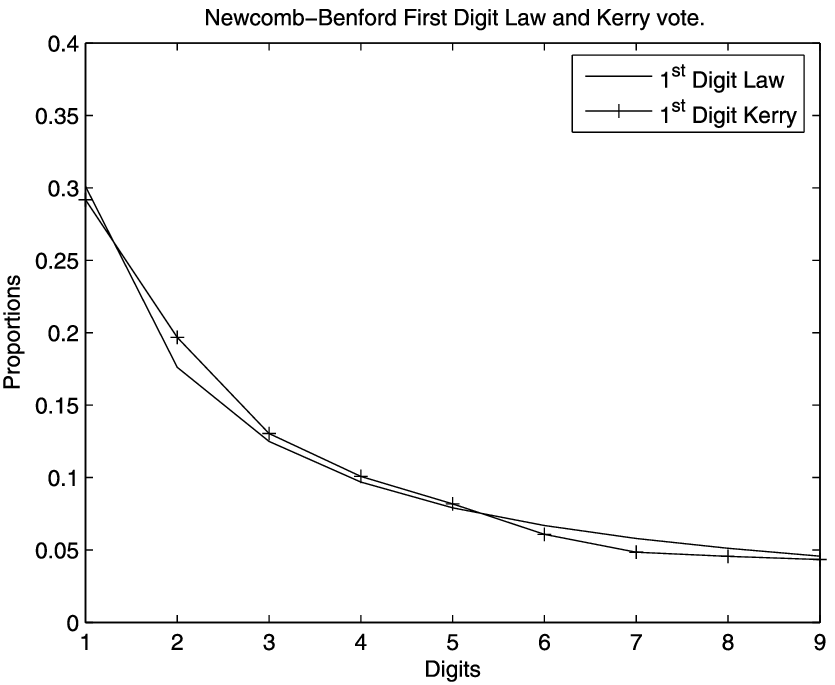}
\caption{Kerry's digit proportions vs. N--B Law for the $1$st digit.}\label{USAKERYNB1}
\vspace*{-6pt}
\end{figure}

\begin{figure}
\includegraphics[scale=0.97]{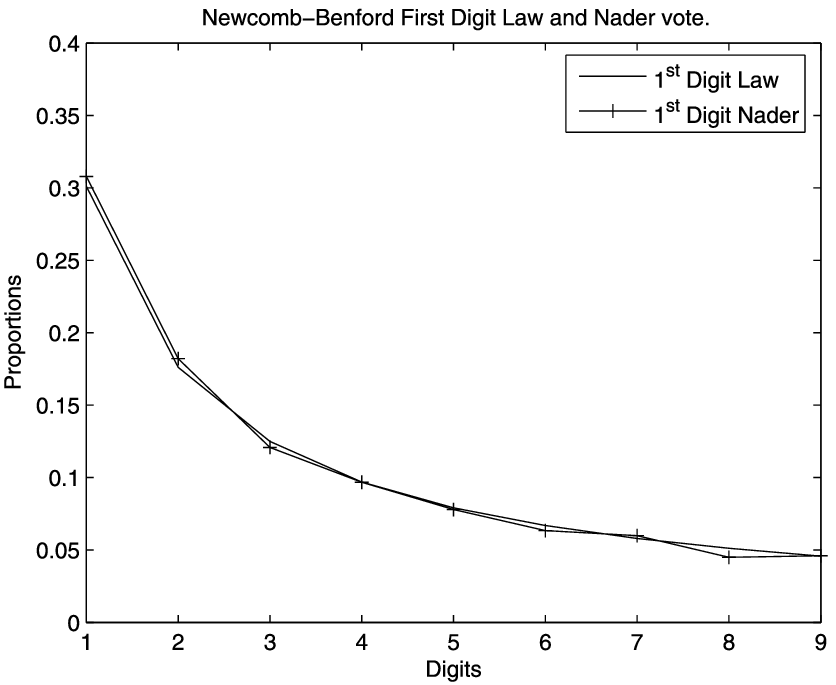}
\caption{Nader's digit proportions vs N--B Law for the $1$st digit.}\label{USANADERNB1}
\label{T5:USNB2}
\vspace*{-2pt}
\end{figure}

\begin{figure}

\includegraphics[scale=0.97]{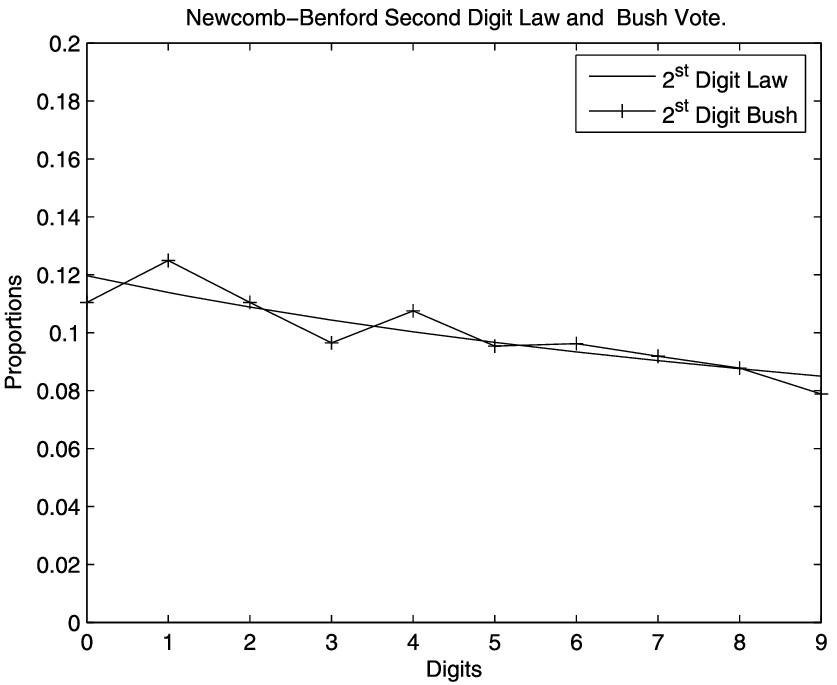}

\caption{Bush's digit proportions vs. N--B Law for the $2$nd digit.}\label{USABUSHNB2}
\vspace*{-1pt}
\end{figure}

\begin{figure}

\includegraphics[scale=0.97]{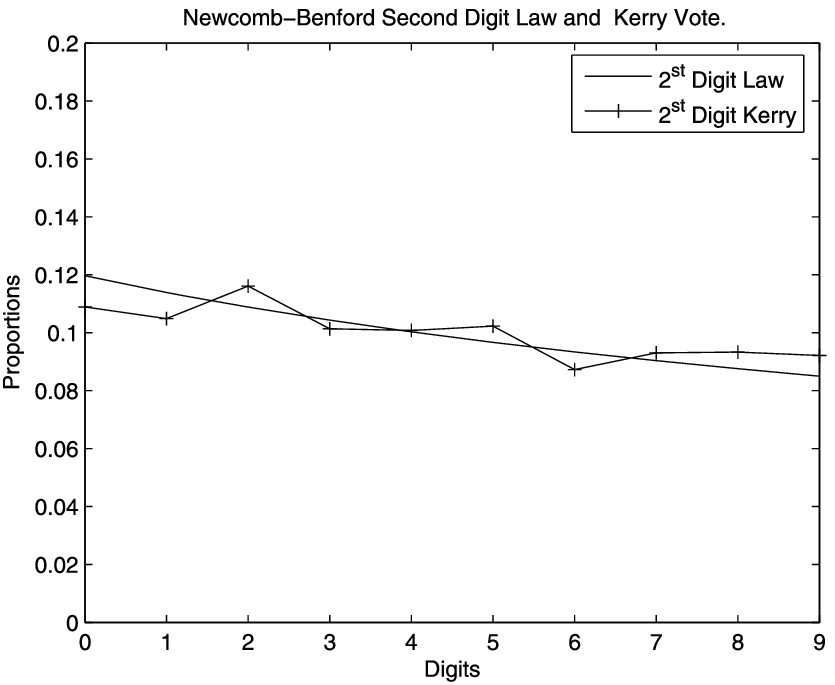}%
\caption{Kerry's digit proportions vs. N--B Law for the $2$nd
digit.}\label{USAKERRYNB2}
\vspace*{-6pt}
\end{figure}

\begin{figure}

\includegraphics[scale=0.97]{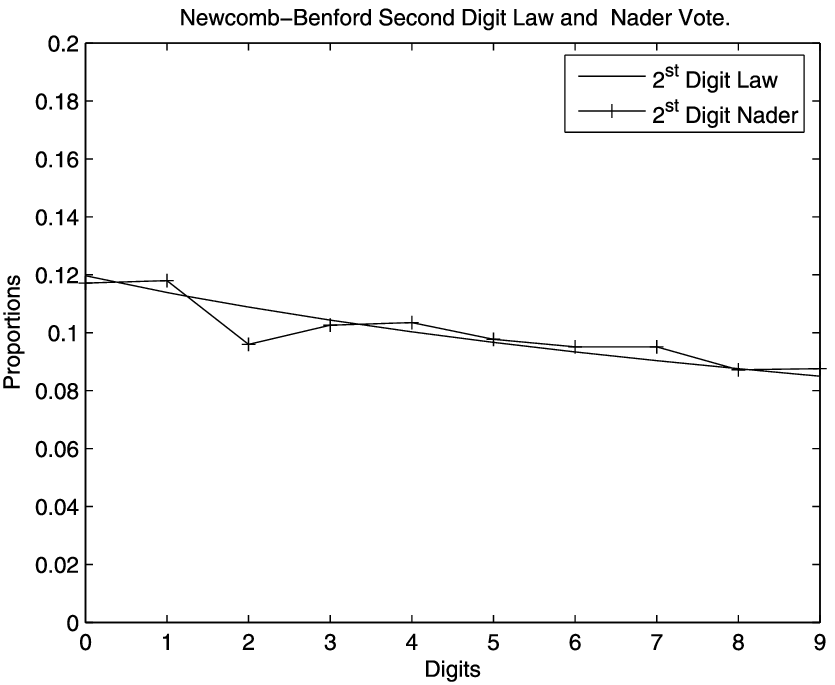}%
\caption{Nader's digit proportions vs. N--B
Law for the $2$nd digit.}
\label{USANADERNB2}
\vspace*{-3pt}
\end{figure}

The first case in point is the 2004 USA
presidential election, Tables~\ref{T:6} and \ref{T:7} and Figures~\ref{USAjoint}--\ref{USANADERNB2}. The data at the
level of counties can be found at
\texttt{\href{http://us.cnn.com/ELECTION/2004/pages/results/}{http://}
\href{http://us.cnn.com/ELECTION/2004/pages/results/}{us.cnn.com/ELECTION/2004/pages/results/}}.
%
%
%
%
%
%
%
%

%
%
%
%
%
%

(Note: Nader's votes had to be constructed from alternative
sources.) This is one of the best case studies we know about the
inadequacy of $p$-values when compared to the impressive fit of the
NBL with both the first and the second digit, and even with the
joint density of first and second digit. For example, in the case of
Bush's votes, for the first digit the fit is excellent, but the
$p$-value is only $0.003$, significant even at $0.01$ level. On the
other hand, the absolute minimum of posterior probabilities of the
null hypothesis is $0.05$, over sixteen times the $p$-value.\vadjust{\goodbreak} Note that
this is only a lower bound over all possible prior distributions,
which is certainly understating the true evidence. Not surprisingly,
a real Bayes Factor leads to a posterior
probability of almost one.

The best fit is Nader's votes, which is not significant, neither for
the first or the second digit NBL, and so not surprisingly, the
posterior probabilities of compliance with NBL is one. Bush's and
Kerry's votes first digit tests are significant with small $p$-values,
but the posterior probabilities are virtually one. For the second
digit law the fit in all these cases is excellent. This is
illustrated by Figures~\ref{USAjoint}--\ref{USANADERNB2}.

\subsection{Puerto Rico}

Here we show the data for the three main parties (PNP, PPD and PIP)
in the 1996, 2000 and 2004 elections for governor. The data can be
found at
\url{http://electionspuertorico.org/datos/2004}\break
and
\texttt{\href{http://www.ceepur.org/elecciones2000/escrutinio/datos/}{http://www.ceepur.org/elecciones2000/}}.

The results about the first digit are significant. Moreover, the
posterior probabilities also reject the NBL1. The restricted NBL for
the first digit does not show a big improvement either. This may be
due to the fact that in electoral processes, the upper bounds (the
total number of electors per polling station) is not typically fixed
across the population of polling stations. However, the second digit
shows an excellent fit to the NBL2 Law, and the results with
restrictions do not change much, illustrating again that the effect
of bounds in the second digit is usually smaller than for the first
digit NBL.

\subsection{Venezuela}

\subsubsection{Referendum}

The 2004 {\spaceskip=0.2em plus 0.05em minus 0.02emPresidential Revoca\-tory Referendum in
Venezuela has attracted con\-sid\-er\-able interest and controversy. (Data
from the Refe-\break rendum can be found at}
\texttt{\href{http://www.cne.gob.ve/referendum_presidencial2004/}{http://www.cne.gob.ve}},\break
\url{http://www.venezuela-referendum.com},
\texttt{\href{https://sites.google.com/a/upr.edu/probability-and-statistics/data-files-1}%
{https://}
\href{https://sites.google.com/a/upr.edu/probability-and-statistics/data-files-1}%
{sites.google.com/a/upr.edu/probability-and-}
\href{https://sites.google.com/a/upr.edu/probability-and-statistics/data-files-1}%
{statistics/data-files-1}},
\texttt{\href{http://esdata.info/2004}{http://esdata.info/}
\href{http://esdata.info/2004}{2004}}.)

One of the most interesting features of this process is that
it was partly manual and partly electronic, with the majority of the
polling stations having electronic voting, but a sizeable proportion
being manual. Here, NO means in favor of the President and SI
against.

The most salient feature is that the electronic NO votes
give evidence against NB2 Law. Figure~\ref{VENE04:1} shows that the
second digits seem to be Uniformly distributed. This is not the case
for manual votes, or for the SI electronic votes. This finding is
quite informative: \textit{The electronic votes in favor of the
government need closer scrutiny}.

%
%
\begin{table}
\tabcolsep=5pt
\tablewidth=\columnwidth
\caption{Results of the 1996 Governor Elections in Puerto Rico}\label{T:8}
\begin{tabular*}{\columnwidth}{@{\extracolsep{\fill}}lcccc@{}}
\hline
\textbf{Puerto Rico}
\\
\multicolumn{1}{@{}c}{\textbf{1996}}
&$\bolds{m}$
&$\bolds{P(H_0|\mathit{data})}$
&$\bolds{p}$\textbf{-values}
&$\bolds{\underline{P}(H_0|\mathit{data})}$\\\hline
NB2 PNP&1836&1.000&0.554&$>0.5$\phantom{00} \\
NB2 PPD&1839&1.000&0.138&\phantom{0.}0.426\\
NB2 PIP&1466&1.000&0.104&\phantom{0.}0.390\\\hline
\end{tabular*}
\vspace*{-7pt}
\end{table}

%
%
\begin{table}
\tabcolsep=5pt
\tablewidth=\columnwidth
\caption{Results of the 2000 Governor Elections in Puerto Rico}\label{T:9}
\begin{tabular*}{\columnwidth}{@{\extracolsep{\fill}}lcccc@{}}
\hline
\textbf{Puerto Rico}
\\
\multicolumn{1}{@{}c}{\textbf{2000}}
&$\bolds{m}$
&$\bolds{P(H_0|\mathit{data})}$
&$\bolds{p}$\textbf{-values}
&$\bolds{\underline{P}(H_0|\mathit{data})}$
\\\hline
NB2 PNP&1823&1.000&0.979& $ > 0.5 $ \\
NB2 PPD&1878&1.000&0.436& $ > 0.5 $ \\
NB2 PIP&1579&1.000&0.450& $ > 0.5 $\\\hline
\end{tabular*}
\end{table}
%
%
%
\begin{table}
\tabcolsep=5pt
\tablewidth=\columnwidth
\caption{Results of the 2004 Governor Elections in Puerto Rico}\label{T:10}
\begin{tabular*}{\columnwidth}{@{\extracolsep{\fill}}lcccc@{}}
\hline
\textbf{Puerto Rico}
\\
\multicolumn{1}{@{}c}{\textbf{2004}}
&$\bolds{m}$
&$\bolds{P(H_0|\mathit{data})}$
&$\bolds{p}$\textbf{-values}
&$\bolds{\underline{P}(H_0|\mathit{data})}$\\\hline
NB2 PPD&1924&1.000&0.154& \phantom{$>0.$ }0.440\phantom{00.}\\
NB2 PND&1917&1.000&0.538& $>0.5$\phantom{00.}\\
NB2 PIP&1402&1.000&0.822& $>0.5$\phantom{00.} \\\hline
\end{tabular*}
\end{table}

\subsubsection{Venezuela 2000}

For comparison purposes\break the~Venezuelan presidential
election of 2000 (the presidential election previous to the recall
referendum of 2004) is presented here. (Data can be found~in:
\texttt{\href{https://sites.google.com/a/upr.edu/probability-and-statistics/data-files-1}%
{https://sites.google.com/a/upr.edu/probability-}\break
\href{https://sites.google.com/a/upr.edu/probability-and-statistics/data-files-1}%
{and-statistics/data-files-1}},
\texttt{\href{http://esdata.info/downloads/ELECCIONES2000.zip}%
{http://esdata.}\break
\href{http://esdata.info/downloads/ELECCIONES2000.zip}%
{info/downloads/ELECCIONES2000.zip}}.)

%
%
\begin{figure}
\centering
\begin{tabular}{@{}cc@{}}

\includegraphics[scale=0.97]{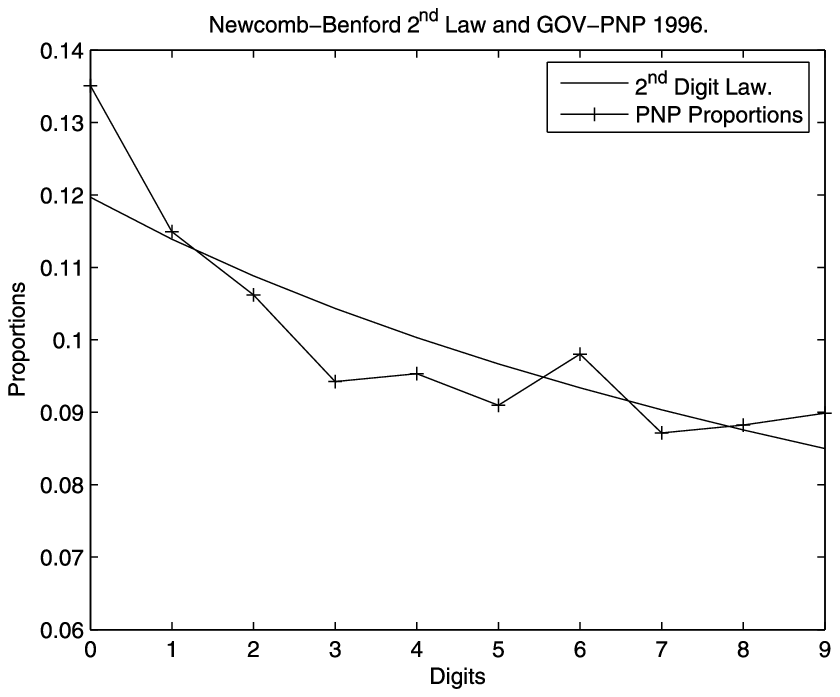}
\\
\footnotesize{(a) Puerto Rico Elections 1996 PNP Party.}\\[3pt]

\includegraphics[scale=0.97]{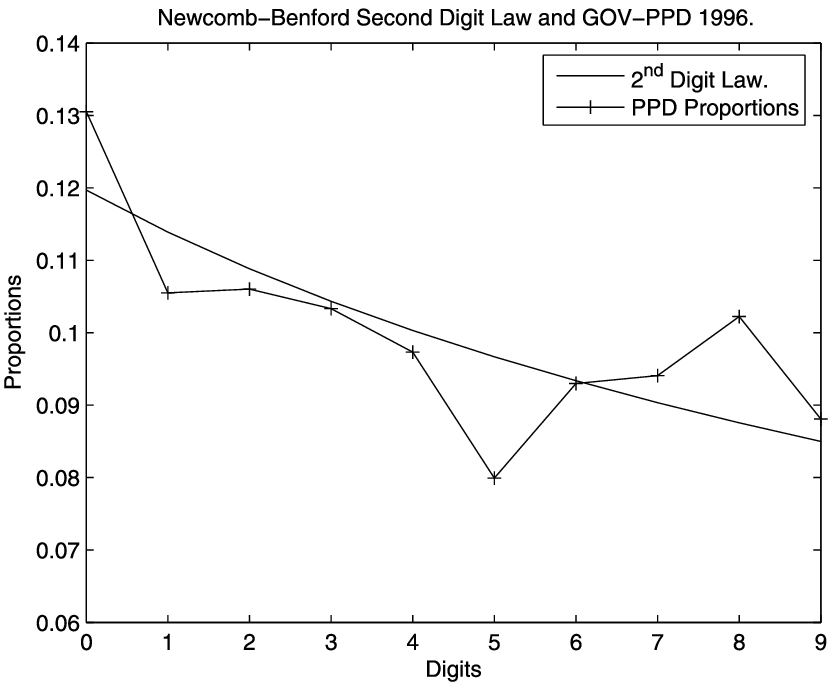}
\\
\footnotesize{(b) Puerto Rico Elections 1996 PPD Party.} \\[3pt]

\includegraphics[scale=0.97]{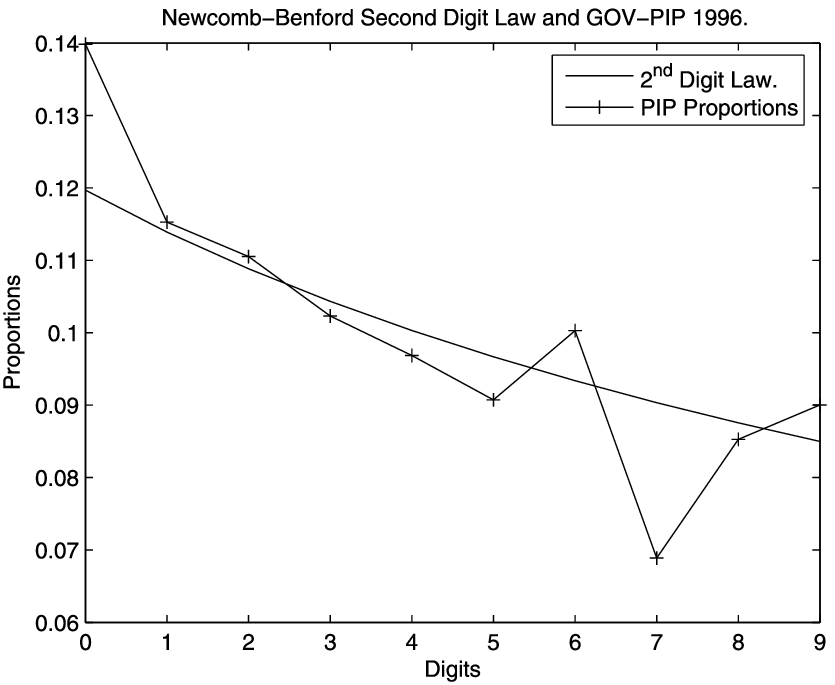}
\\
\footnotesize{(c) Puerto Rico Elections 1996 PIP Party.}
\end{tabular}
\caption{Puerto Rico 1996 Elections compared with the Newcomb--Benford
Law for the second digit.} \label{PR1996}
\end{figure}

%
%
\begin{figure}
\centering
\begin{tabular}{@{}c@{}}

\includegraphics[scale=0.97]{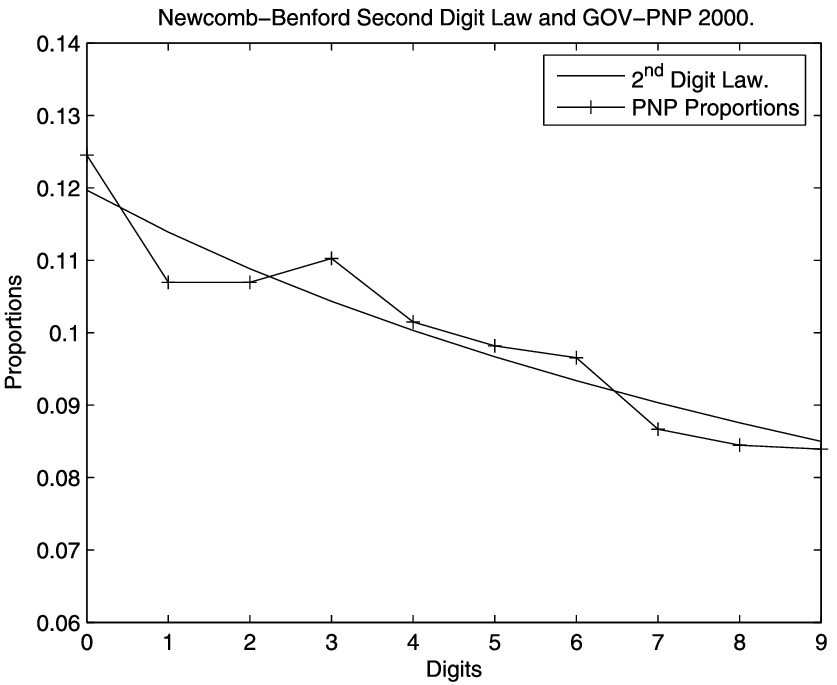}
\\
\footnotesize{(a) Puerto Rico Elections 2000 PNP Party.}\\[3pt]

\includegraphics[scale=0.97]{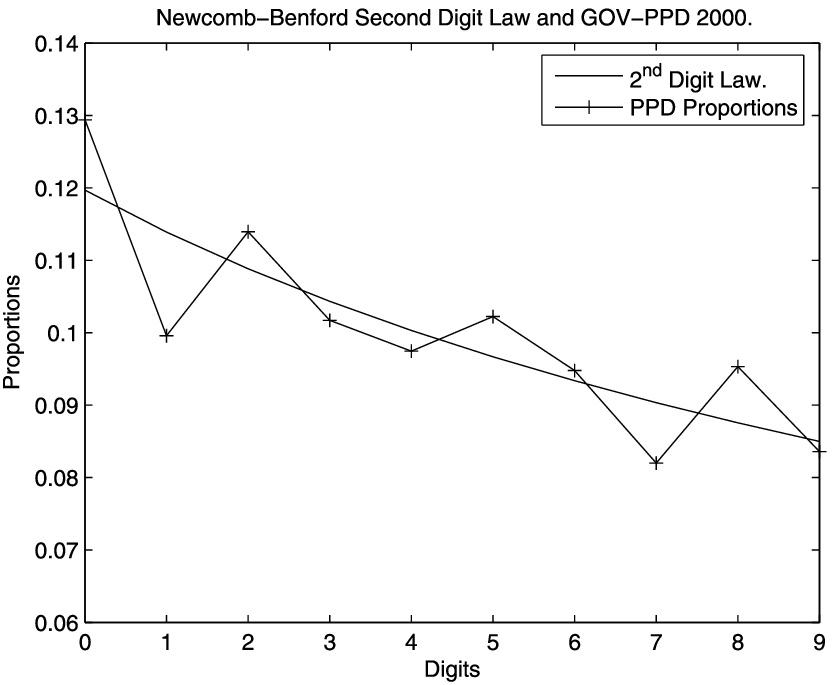}
\\
\footnotesize{(b) Puerto Rico Elections 2000 PPD Party.}\\[3pt]

\includegraphics[scale=0.97]{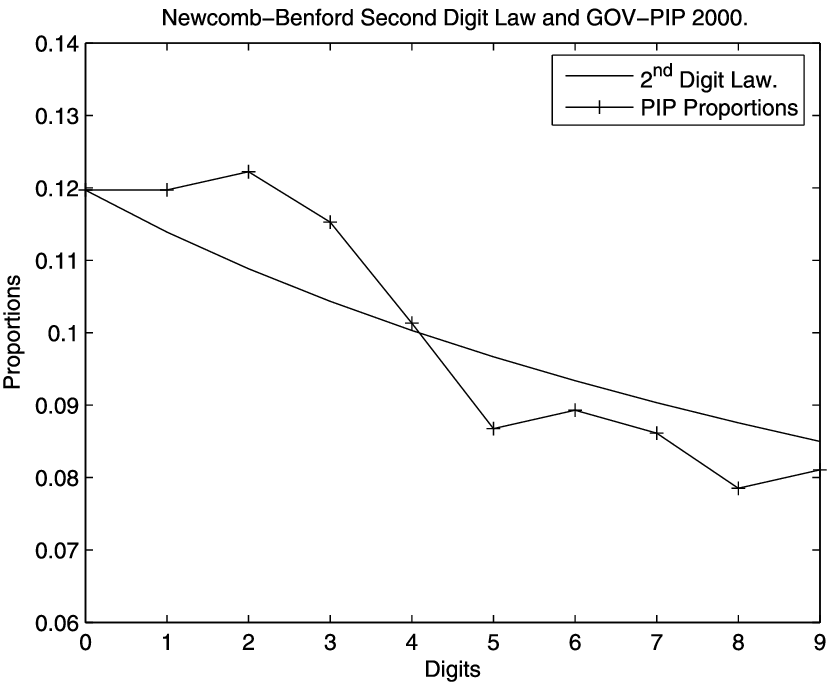}
\\
\footnotesize{(c) Puerto Rico Elections 2000 PIP Party.}
\end{tabular}
\caption{Puerto Rico 2000 Elections compared with the Newcomb--Benford
Law for the second digit.} \label{PR2000}
\end{figure}

%
%
\begin{figure}
\centering
\begin{tabular}{@{}cc@{}}

\includegraphics[scale=0.97]{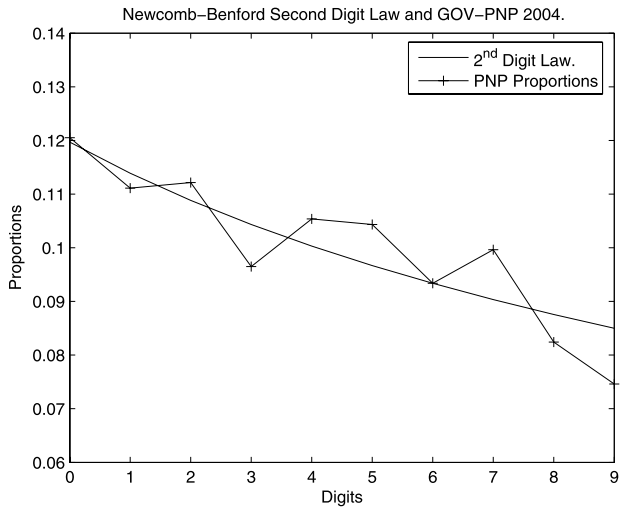}
\\
\footnotesize{(a) Puerto Rico Elections 2004 PNP Party.}\\[3pt]

\includegraphics[scale=0.97]{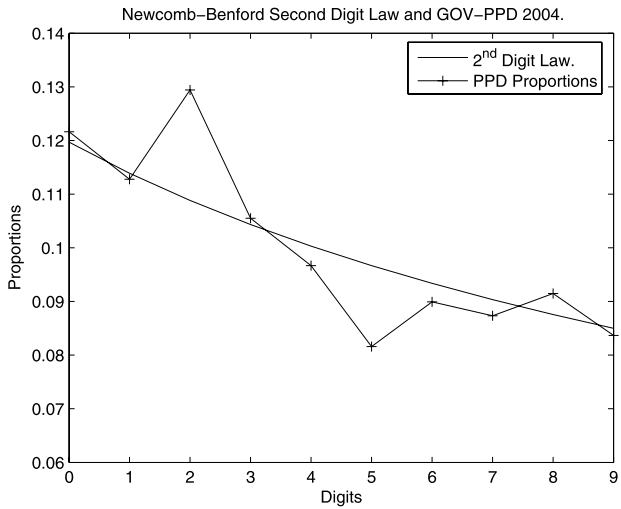}
\\
\footnotesize{(b) Puerto Rico Elections 2004 PPD Party.}\\[3pt]

\includegraphics[scale=0.97]{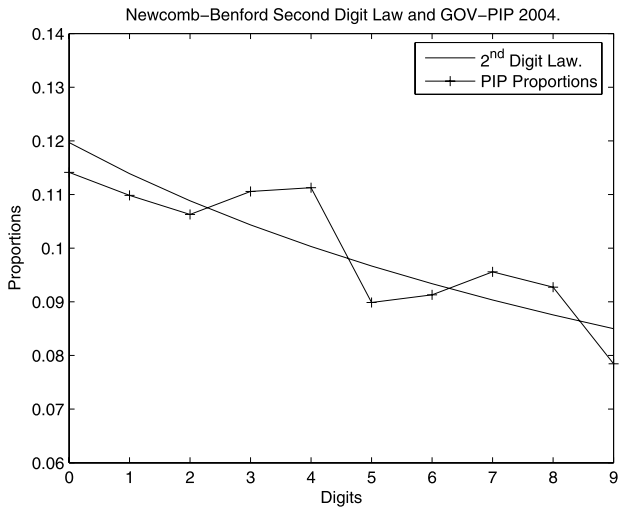}
\\
\footnotesize{(c) Puerto Rico Elections 2004 PIP Party.}
\end{tabular}
\caption{Puerto Rico 2004 Elections compared with the Newcomb--Benford
Law for the first digit. } \label{PR2004}
\end{figure}

%
\begin{figure}

\includegraphics{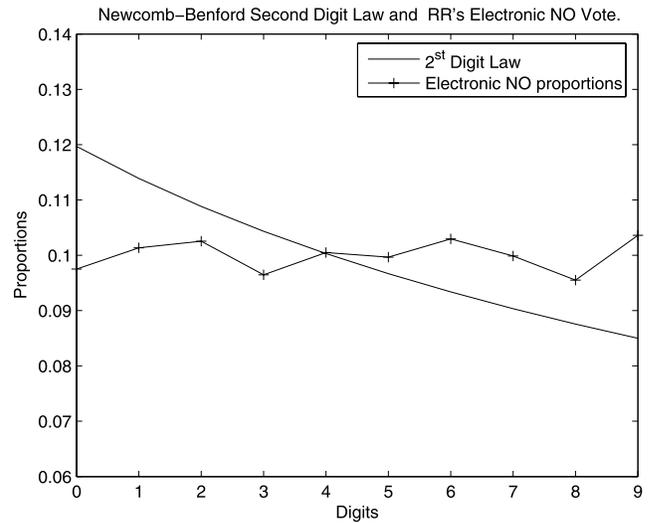}

\caption{Venezuela Revocatory Referendum Electronic NO Votes proportions.
Venezuela Revocatory Referendum Electronic Votes Proportions compared with
the Newcomb--Benford Law's proportions for Second digit. This is the
only compelling rejection of
the NBL2 law.}
\label{VENE04:1}
\vspace*{-6pt}
\end{figure}
%
%
\begin{figure}

\includegraphics{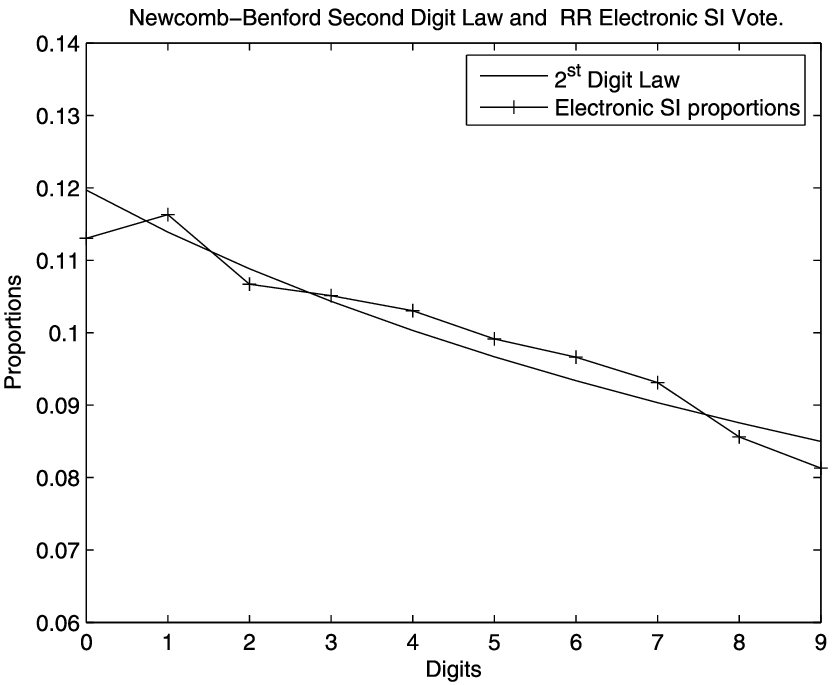}

\caption{Venezuela Revocatory Referendum Electronic SI Votes proportions.
Venezuela Revocatory Referendum Electronic Votes Proportions compared with
the Newcomb--Benford Law's proportions for Second digit. }
\label{VENE04:2}
\end{figure}

%
\begin{figure}

\includegraphics{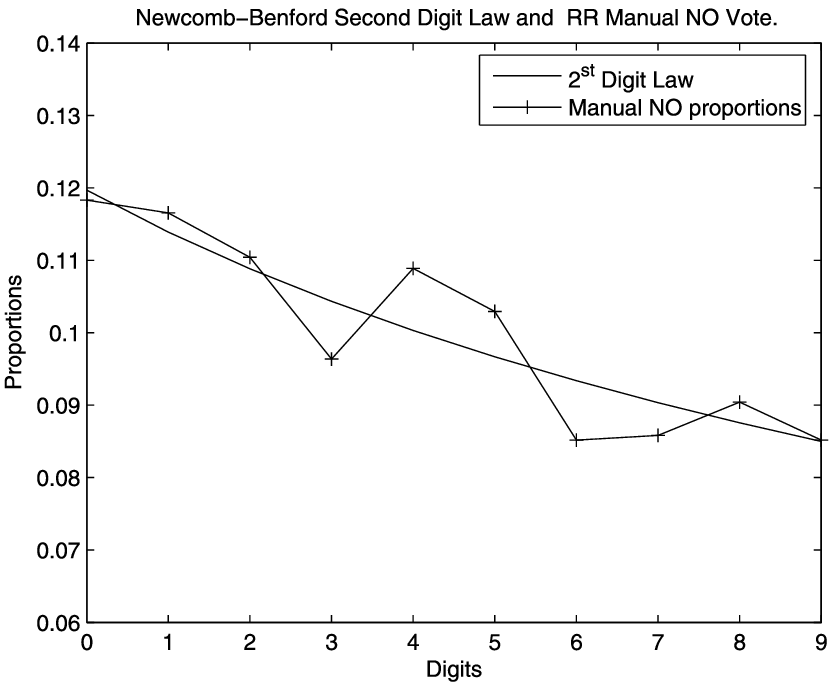}%
\vspace*{-3pt}
\caption{Venezuela Revocatory Referendum Electronic NO Votes proportions.
Venezuela Revocatory Referendum Manual Votes Proportions compared with
the Newcomb--Benford Law's proportions for Second digit. }
\label{VENE04:3}
\vspace*{-6pt}
\end{figure}

%
\begin{figure}

\includegraphics{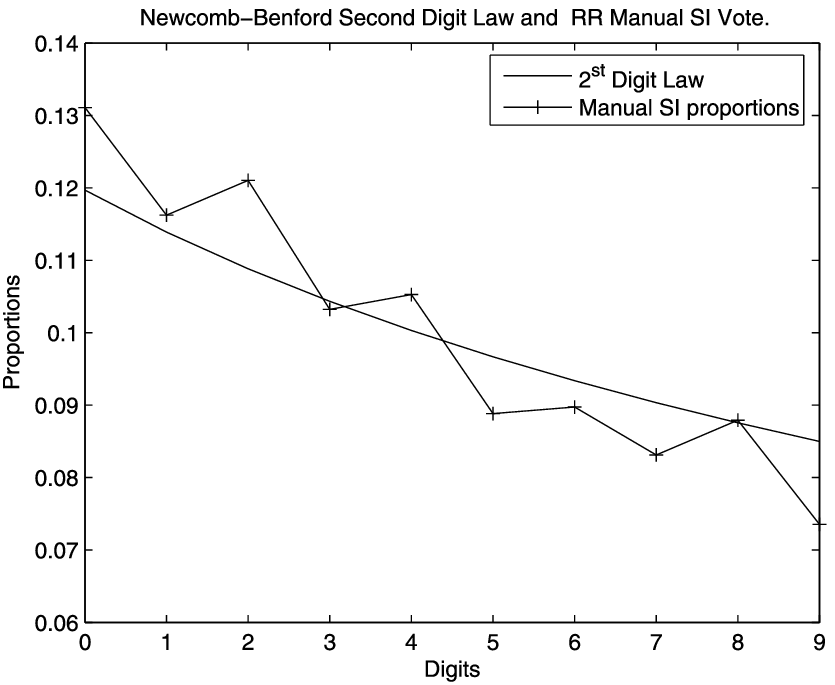}%
\vspace*{-3pt}
\caption{Venezuela Revocatory Referendum Electronic SI Votes proportions.
Venezuela Revocatory Referendum Manual Votes Proportions compared with
the Newcomb--Benford Law's proportions for Second digit. }
\label{VENE04:4}
\vspace*{-4pt}
\end{figure}

%
\begin{figure}

\includegraphics{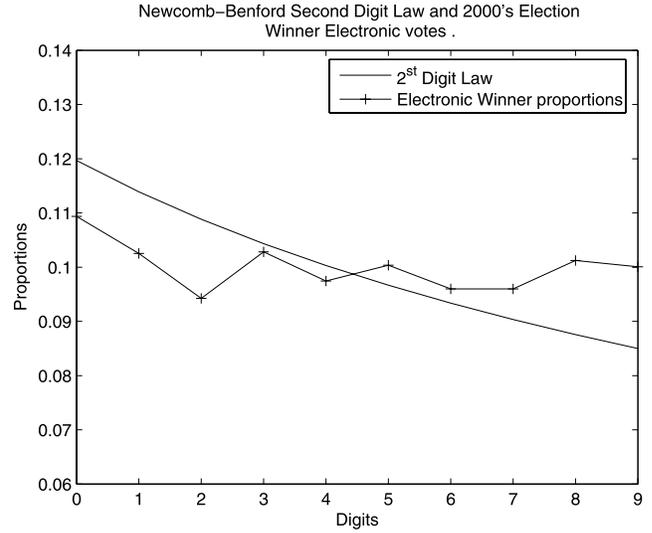}%
\vspace*{-3pt}
\caption{Venezuela 2000 Election Electronic Votes in favor of the
Winner compares with
Newcomb--Benford Law's proportions for Second digit.}
\label{VENE00:1}
\end{figure}

%
\begin{figure}

\includegraphics{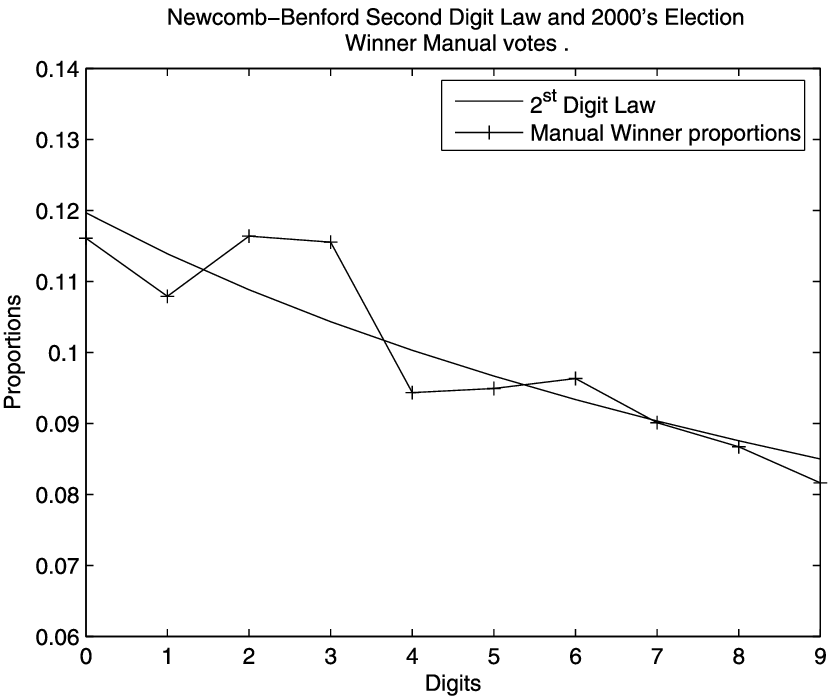}%
\vspace*{-3pt}
\caption{Venezuela 2000 Election Manual Votes proportions in favor of
Winner compares with
Newcomb--Benford Law's proportions for Second digit. }
\label{VENE00:2}
\end{figure}

Here none of the candidates for either manual
or electronic show compelling evidence against the NB2 Law, although
the winning electronic voting results in a posterior probability
smaller than the others. Although to a lesser extent than in the
2004 referendum, this result may indicate the need for a~closer
scrutiny of the winning electronic votes.

\section{Conclusions}\label{sec:5} The main conclusions to be reached
here are as follows:
\begin{enumerate}
\item At a technical level: (i) the RNBL is a substantial
generalization of the NBL that enlarges its domain of applications.
However, in the electoral processes presented here, the differences
in the results with and without the restriction did not change much.
This may be due to the fact that there is no constant upper bound,
since the total number of electors is not the same for all polling
stations. However, it is the case that the second digit law is far
less affected by restrictions than the first digit law. (ii) The
second digit NBL2 is a useful test for quick detection of anomalous
behavior in electronic or manual elections. (iii)~The Universal Lower
Bound and even more so, Bayes Factors, are appropriate measures of
evidence of the fit to the law, and $p$-values are not, particularly
for large data sets like the electoral data.

%
\begin{table*}[t]
\caption{Results of the 2004 Presidential Recall Referendum in
Venezuela for electronic votes}\label{T:11}
\begin{tabular*}{\textwidth}{@{\extracolsep{\fill}}lccccc@{}}
\hline
\textbf{Venezuela RR}
&$\bolds{m}$
&\textbf{Median}
&$\bolds{P(H_0|\mathit{data})}$
&$\bolds{p}$\textbf{-values}
&$\bolds{\underline{P}(H_0|\mathit{data})}$\\\hline
No Electronic NB2&19064&263&0.000&0.000&0.000\\
Si Electronic NB2&19063&172&1.000&0.024&0.196\\\hline
\end{tabular*}
\end{table*}
%

\begin{table*}[t]
\caption{Results of the 2004 Presidential Recall Referendum in
Venezuela for Manual votes}\label{T:12}
\begin{tabular*}{\textwidth}{@{\extracolsep{\fill}}lcd{3.0}ccc@{}}
\hline
\textbf{Venezuela RR}
&$\bolds{m}$
&\textbf{Median}
&$\bolds{P(H_0|\mathit{data})}$
&$\bolds{p}$\textbf{-values}
&$\bolds{\underline{P}(H_0|\mathit{data})}$\\\hline
No Manual NB2&4556&190&1.000&0.155&0.440\\
Si Manual NB2&4379&76&1.000&0.003&0.047\\\hline
\end{tabular*}
\end{table*}

\begin{table*}[t]
\caption{Results of the 2000 Election in Venezuela for electronic
votes}\label{T:13}
\begin{tabular*}{\textwidth}{@{\extracolsep{\fill}}lccccc@{}}
\hline
\textbf{Venezuela 2000}
&$\bolds{m}$
&\textbf{Median}
&$\bolds{P(H_0|\mathit{data})}$
&$\bolds{p}$\textbf{-values}
&$\bolds{\underline{P}(H_0|\mathit{data})}$\\\hline
Winner Electronic NB2&6876&486&0.129&0.000&0.000\\
Runner up Electronic NB2&6872&265&1.000&0.017&0.160\\\hline
\end{tabular*}
\end{table*}
%
\begin{table*}[t]
\caption{Results of the 2000 Election in Venezuela for manual votes}\label{T:14}
\begin{tabular*}{\textwidth}{@{\extracolsep{\fill}}lcd{3.0}ccc@{}}
\hline
\textbf{Venezuela 2000}
&$\bolds{m}$
&\textbf{Median}
&$\bolds{P(H_0|\mathit{data})}$
&$\bolds{p}$\textbf{-values}
&$\bolds{\underline{P}(H_0|\mathit{data})}$\\\hline
Winner Manual NB2&3540&103&1.000&0.366&\multicolumn{1}{c}{\hspace*{-3pt}$>0.5$}\\
Runner up Manual NB2&3219&52&1.000&0.006&\phantom{00}0.081\\\hline
\end{tabular*}
\end{table*}

%
\begin{figure}

\includegraphics{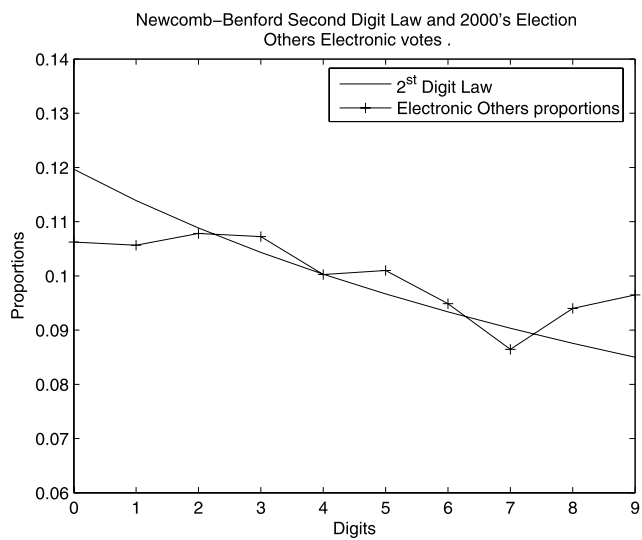}

\caption{Venezuela 2000 Election Electronic Votes proportions of the
Loser compares with
Newcomb--Benford Law's proportions for Second digit. }
\label{VENE00:3}
\end{figure}

\begin{figure}

\includegraphics{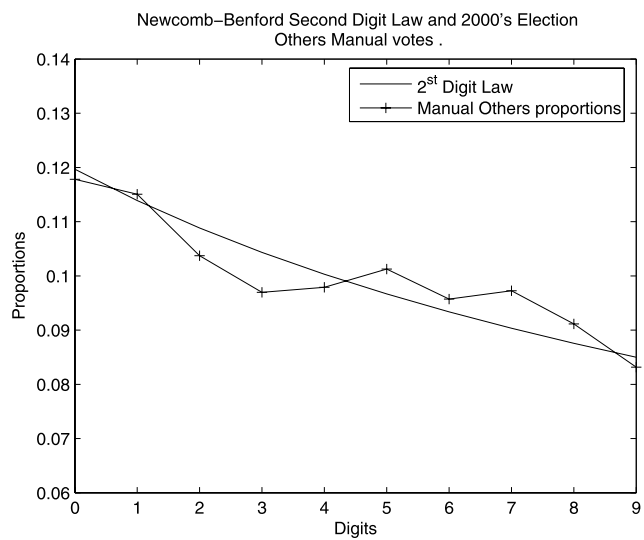}

\caption{Venezuela 2000 Election Manual Votes proportions of the Loser
compares with
Newcomb--Benford Law's proportions for Second digit.}\label{VENE00:4}
\end{figure}

\item Regarding the detection of anomalies: (i) the USA 2004
elections show a remarkable fit to the first digit Newcomb--Benford
Law, and also to the second digit NBL. All the manual elections show
support for the second digit\vadjust{\goodbreak} NBL law. (ii) On the other hand, the
electronic results of the votes in favor of the NO in the Recall
Referendum violate the NB2 law. This is surprising, since the
manual votes in favor and against, as well as the electronic votes
in favor of the opposition, fit the law reasonably well. In the
previous 2000 Venezuelan presidential elections, there is no
compelling evidence against the law, although again the electronic
results in favor of the winner show only about $13\%$ of posterior
probability in favor of the law.
\end{enumerate}

Our methods, particularly the use of the Second Digit
Newcomb--Benford Law, add to the increasing literature on measures
of surprise and legitimate suspicion on electoral processes,
particularly but not restricted to electronic voting. The NBL2,
since our original suggestion in 2004, is becoming a standard tool
on what has been termed by Mebane as ``Election Forensics.''

\section*{Acknowledgments}

NSF Grants 0604896 and 0630927 gave partial support for this research.
LP acknowledges the invitation by the
Faculty Association of the
Universidad Sim\'{o}n Bol\'{\i}var, Caracas, to present the first draft
in 2004. A~detailed and constructive report by
a referee and Associate Editor
helped us improve the presentation. We also thank our colleagues M. E.
P\'{e}rez and P. Rodr\'{\i}guez-Esquerdo for
very useful suggestions. Finally, we are most grateful to the Carter Center
for giving publicity to our unpublished draft and workshop presentation.


\begin{thebibliography}{99}

\bibitem[\protect\citeauthoryear{Benford}{1938}]{Benford}
\textsc{Benford, F.} (1938).
The law of anomalous numbers.
\textit{Proc. Amer. Philos. Soc.} \textbf{78} 551--572.

\bibitem[\protect\citeauthoryear{Berger and Pericchi}{2001}]{BP01}
\textsc{Berger, J. O.} and \textsc{Pericchi, L. R. }(2001).
Objective Bayesian methods for model selection:
Introduction and comparison (with discussion).
In \textit{Model Selection} 135--207. IMS, Beachwood,
OH.
\MR{2000753}

\bibitem[\protect\citeauthoryear{Buttorff}{2008}]{Buttorff}
\textsc{Buttorff, G.} (2008).
Detecting fraud in America's gilded age.
Technical report, Univ. Iowa.

\bibitem[\protect\citeauthoryear{The Carter Center} {2005}]{Cartercenter05}
\textsc{The Carter Center} (2005).
Observing the Venezuela Presidential Recall Referendum.
\texttt{%
\href{http://www.cartercenter.org/documents/2020.pdf}%
{http://www.cartercenter.}
\href{http://www.cartercenter.org/documents/2020.pdf}%
{org/documents/2020.pdf}}
Comprehensive Report. Feb. 2005.

\bibitem[\protect\citeauthoryear{The Economist (US)}{2007}]{Economist}
The Economist. Feb. 24th--March 2nd, 2007. Pages 93--94.
Political Science: Election forensics.
\texttt{%
\href{http://www.economist.com/science/displaystory.cfm?story_id=8733747}{http://www.economist.}
\href{http://www.economist.com/science/displaystory.cfm?story_id=8733747}{com/science/}}.

\bibitem[\protect\citeauthoryear{Hill}{1995}]{THill95a}
\textsc{Hill, T.} (1995).
Base-invariance implies Benford's law.
\textit{Proc. Amer. Math. Soc.} \textbf{123} 887--895.
\MR{1233974}

\bibitem[\protect\citeauthoryear{Hill}{1996}]{THill96}
\textsc{Hill, T.} (1996).
A statistical derivation of the Significant-Digit Law.
\textit{Statist. Sci.} \textbf{10} 354--363.
\MR{1421567}

\bibitem[\protect\citeauthoryear{Mebane}{2006}]{Mebane06}
\textsc{Mebane, W. R.} (2006).
Election Forensics: The Second-digit Benford's Law
Test and Recent American Presidential Elections.
Election Fraud Conference, Salt Lake Ciy, Utah, September 29--30.

\bibitem[\protect\citeauthoryear{Mebane}{2007a}]{Mebane07a}
\textsc{Mebane, W. R.} (2007a).
Statistics for Digits. 2007 Summer Meeting of the Political Methodology Society,
Pennsylvania State Univ., July 18--21.

\bibitem[\protect\citeauthoryear{Mebane}{2007b}]{Mebane07b}
\textsc{Mebane, W. R.} (2007b).
Evaluating voting systems to improve and verify accuracy.
Presented at the Annual Meeting of the American.
Association for the Advancement of Science, San Francisco, Feb. 16, 2007.
Available at
\texttt{\href{http://em.fis.unam.mx/\textasciitilde mochan/elecciones/paperMebane.pdf}%
{http://}
\href{http://em.fis.unam.mx/\textasciitilde mochan/elecciones/paperMebane.pdf}%
{em.fis.unam.mx/\textasciitilde mochan/elecciones/paperMebane.pdf}}.


\bibitem[\protect\citeauthoryear{Newcomb}{1881}]{NewS}
\textsc{Newcomb S.} (1881).
Note on the frequency of use of the {Different Digits} in {Natural Numbers}.
\textit{Amer. J. Math.} \textbf{4} 39--40.
\MR{1505286}
\

\bibitem[\protect\citeauthoryear{Nigrini}{1995}]{Nigrini95}
\textsc{Nigrini, M.} (1995).
A taxpayer compliance application of Benford's Law.
\textit{J. Amer. Taxation Assoc.} \textbf{18} 72--91.


\bibitem[\protect\citeauthoryear{Pericchi and Torres}{2004}]{PeriTor04}
\textsc{Pericchi, L. R.} and \textsc{Torres, D.} (2004).
La Ley de Newcomb--Benford y sus aplicaciones al Referendum Revocatorio
en Venezuela. Reporte T\'{e}cnico no-definitivo 2a,
Octubre 01, 2004. Presented on Sept. 23, 2004 on the Third Universidad
Simon Bolivar Seminar on:
Statistical Analyses of the Venezuelan Recall Referendum.
Available at
\texttt{%
\href{https://sites.google.com/a/upr.edu/probability-and-statistics/home/techical-reports}%
{https://sites.google.com/a/upr.edu/probability-}
\href{https://sites.google.com/a/upr.edu/probability-and-statistics/home/techical-reports}%
{and-statistics/home/techical-reports}}.

\bibitem[\protect\citeauthoryear{Pietronero, Tosatti and Vespignani}{2001}]{Piet01}
\textsc{Pietronero, L.}, \textsc{Tosatti, E.} and \textsc{Vespignani, A.} (2001).
Explaining the uneven distribution of numbers in nature: The Laws of
Benford and Zipf.
\textit{Physica A} \textbf{293} 297--304.
\bibitem[\protect\citeauthoryear{Raimi}{1976}]{Raimi}
\textsc{Raimi, R.} (1976).
The first digit problem.
\textit{Amer. Math. Monthly} \textbf{102} 322--327.
\MR{0410850}

\bibitem[\protect\citeauthoryear{Sellke, Bayarri and Berger}{2001}]{Ber}
\textsc{Sellke, T.}, \textsc{Bayarri, M. J.} and \textsc{Berger, O.
J.} (2001).
Calibration of $p$-values for testing precise null hypotheses.
\textit{Amer. Statist.} \textbf{55} 62--71.
\MR{1818723}

\bibitem[\protect\citeauthoryear{Taylor}{2005}]{Taylor2005}
\textsc{Taylor, J.} (2005).
Too many ties? An empirical analysis of the Venezuelan recall
referendum counts.
Technical report.

\bibitem[\protect\citeauthoryear{Taylor}{2009}]{Taylor}
\textsc{Taylor, J.} (2009).
Too many ties? An empirical analysis of the Venezuelan recall
referendum counts. \textit{Statist. Sci.}
To appear.

\bibitem[\protect\citeauthoryear{Torres N\'{u}\~{n}ez}{2006}]{Torres06}
\textsc{Torres N\'{u}\~{n}ez D. A.} (2006).
Newcomb--Benford's Law Applications to
Electoral Processes, Bioinformatics, and the Stock Index.
Supervised by L. R. Pericchi. May 2006. MS thesis.

\bibitem[\protect\citeauthoryear{Torres et al.}{2007}]{TorresJ2007}
\textsc{Torres, J., Fernandez, S., Gamero, A. and Sola, A.} (2007).
How do numbers begin?
(the first digit law). \textit{Eur. J. Phys.} \textbf{28} L17--L25.

\end{thebibliography}
\end{document}